# Automated MRI Field of View Prescription from Region of Interest Prediction by Intra-stack Attention Neural Network

[1]Ke Lei, [1]Ali B.Syed, [2]Xucheng Zhu, [1]John M.Pauly, [1]Shreyas S.Vasanawala
[1]Stanford University, [2]GE Healthcare

# Abstract

## Background

Manual prescription of the field of view (FOV) by MRI technologists is variable and prolongs the scanning process. Often, the FOV is too large or crops critical anatomy.

## Purpose

We propose a deep-learning framework, trained by radiologists' supervision, for predicting region of interest (ROI) and automating FOV prescription.

## Study Type

Retrospective.

## Subjects

284 pelvic and 311 abdominal localizer cases collected between February 2018 and February 2022. 20 pelvic and 20 abdominal cases are used in the test set. There are two series per case, one axial and one coronal.

## Field Strength/Sequence

3.0T scanners. Single shot fast spin echo sequence.

## Assessment

The ROI predicted by the proposed framework is compared with radiologist labeled ROI quantitatively by intersection over union (IoU) and pixel error on boundary positions. It is also assessed qualitatively by a reader study with a radiologist.

## Statistical Tests

We use the t-test for comparing quantitative results from all models and a radiologist. Significance levels of $P<0.05$ and $P<0.01$ are used.

## Results

The proposed model achieves an average IoU of 0.867 and average ROI position error of 9.06 out of 512 pixels on 80 test cases, significantly better (P<0.05) than two baseline models and not significantly different from a radiologist (P>0.12). The FOV prescribed by the proposed framework achieves an acceptance rate of 92% from an experienced radiologist.

## Data Conclusion

Our neural network based ROI prediction model performs comparable to a radiologist. FOV prescription derived from the model predicted ROI (based on MR sampling theory) has a high acceptance rate for clinical application.

At the beginning of clinical MRI exams, a set of localizer images with low spatial resolution and large field of view (FOV) is collected to help define more precise imaging regions for the following diagnostic image acquisitions. MRI technologists select a FOV on the localizer images to plan the next scan. This manual step slows down the scanning workflow, prolonging the overall exam time, and influences image quality. Often technologists are not fully informed of radiologist's requirements for region of interest (ROI). Figure 1 shows two examples of FOVs prescribed by a technologist compared to a radiologist's required FOV. Poor FOV assignments may lead to images with relevant anatomy truncated and a non-diagnostic exam. A conservatively assigned large FOV costs scan time or resolution. Therefore, we propose automating the FOV assignment with models trained by radiologists' ROI labels to get near optimal FOV and streamline scans after the initial localizer.

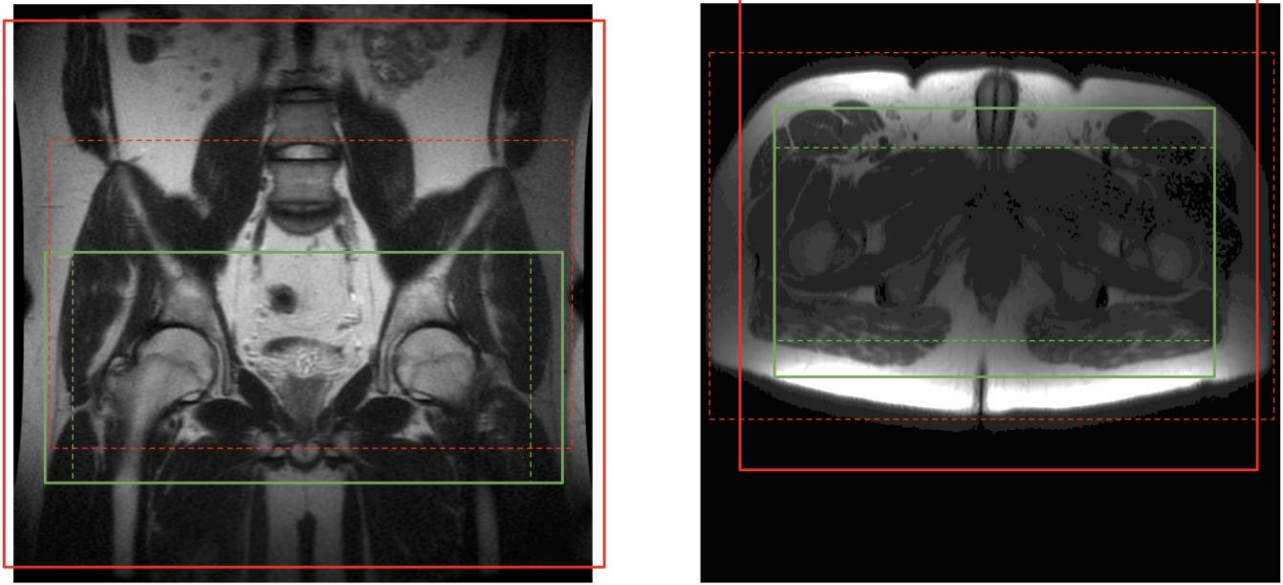

**Figure 1:** Technologist prescribed FOV, and slice range for the orthogonal plane, compared with optimal FOV derived from radiologist's ROI. Red solid boxes represent the FOV prescribed for the same plane as the localizer image. Red dashed boxes indicate the slice range and left-right FOV prescribed for plane orthogonal to that of the localizer image, i.e., axial plane for the coronal localizer and vice versa. The red dashed box in the left sample truncated the bottom part of the ROI. The other three red boxes are much larger than needed.

Studies on brain (1-3) and knee (4) MR images have shown that, for longitudinal studies where precise reproducibility of scan prescription is important, automatic methods achieve less variance than manual prescriptions. (5) presents a brain volume of interest prediction by registration methods that takes nearly one minute. (6-7) present and evaluate an automatic prescription method for knee images based on an anatomical landmark predicted by an active shape model (8) and takes over ten seconds for inference. Deep learning based methods have fast inferences, commonly under one second, and have become popular for organ segmentation tasks. Works on liver (9-10), spleen (11-12), and kidney (13-14) segmentations can be combined to get a single rectangular ROI for the abdomen. However, besides the excess number of models needed, not all ROI can be defined completely from the edges of organs.

We present a framework where the input is a set of localizer images, the output is a rectangular ROI predicted by a convolutional neural network (CNN) based model, and a FOV is then derived from the ROI according to MRI sampling theory. The high-level workflow is illustrated in Figure 2. We compare our model's ROI prediction to that given by two baseline models and by an experienced radiologist, respectively. We compare our model's FOV selection with that of a technologist, and check the clinical acceptability rate of the FOV given by our framework. The purpose of this study is to construct a framework for automating FOV assignment and examine its suitability for clinical application.

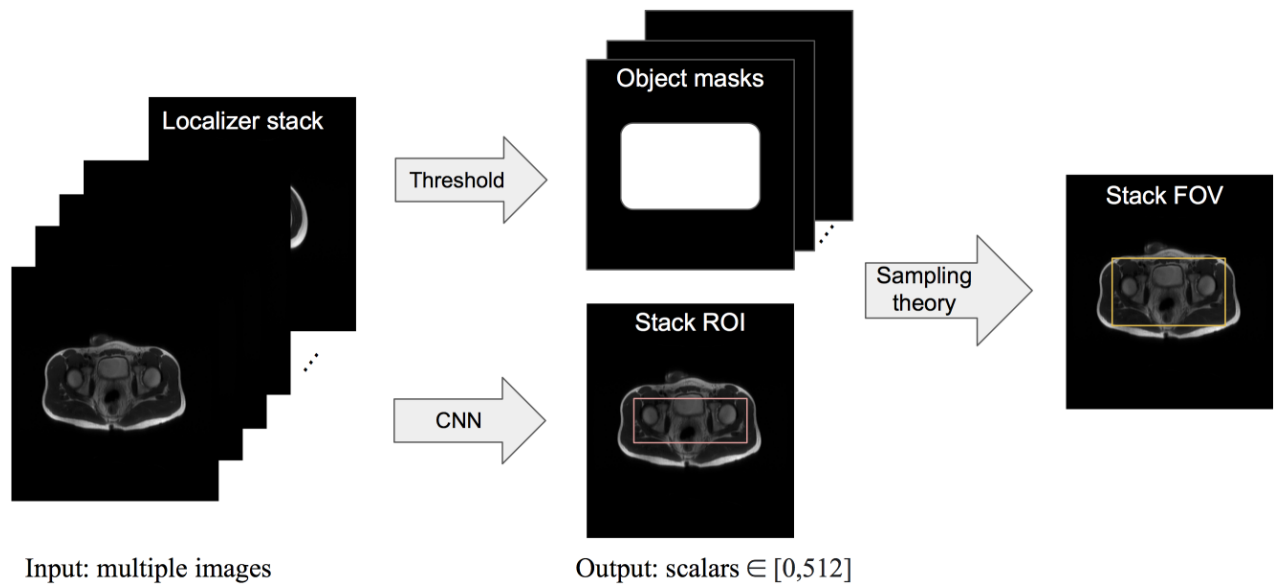

**Figure 2:** Overview of the proposed framework pipeline. First, a CNN based model is trained to predict an ROI for a localizer stack. Then, slice-wise simplified rectangular masks of the object are obtained by setting a threshold on the row or column sums of the pixel values. The final output is the smallest FOV that places aliasing copies of the object masks outside the predicted ROI.

## Materials and Methods

**Dataset Collection**

With IRB approval and informed consent/assent, we retrospectively gathered all axial and coronal single-shot fast spin echo localizer raw data from pelvic or abdominal scans for pediatric patients referred for clinical exams from a four-year period. Sex and age information is not available at data curation stage, but a large range of body size can be observed from the anonymized image samples. We then reconstructed the k-space data using a k-space parallel imaging method. There are 284 pelvic and 311 abdominal cases. The number of slices in a localizer stack ranged from 1 to 40.

Our study uses pediatric data because there is higher variation in body shape from pediatric patients than that from adult patients, making the application more challenging. We chose pelvic and abdominal FOV assignments because these normally base on multiple anatomies and are not organ segmentation problems.

For a given localizer series, one of two radiologists was asked to draw a rectangular box on each stack of localizers representing the desired ROI of subsequent acquisitions. The coordinates of the box's four corners are recorded as the label for each stack. The coronal training sets are labeled by radiologist A (twenty years of MRI experience); the axial training sets are labeled by radiologist B (five years MRI experience). The test sets are labeled by both radiologists independently. The pelvic ROI is labeled based on hip exams, ensuring coverage of the trochanters, symphysis, and hamstring origins. The abdominal ROI is labeled to tightly cover liver, spleen, and kidneys. In abnormal cases, such as peripancreatic fluid collections or kidney transplants in the pelvis, those regions are also covered. Subcutaneous fat is included for coronal scans and excluded for axial scans.

**Data Augmentation**

To augment the training dataset, we first create a horizontal flipped copy of all original image stacks. Then during training, all image stacks are cyclic shifted along the width and height dimensions for an integer number of pixels randomly chosen from sets {-10, -5, 0, 5, 10} and {-20, -10, 0, 10, 20}, respectively. The boundary labels are adjusted accordingly for the augmentation samples.

The input and output of all neural network models in this work are a stack of 2D images and two scalars, respectively. We independently train two instances of each model to output a pair of scalars in the range of 0 and 512, representing the top and bottom, or left and right boundaries of the ROI box. We use the mean-squared error as the training loss for all models. We train two instances of each model to predict the left-right and top-bottom boundaries independently because empirically this results in better performance than using one or four instances.

**Two Baseline Models**

We first present two standard CNN models on end-to-end ROI regression from stacks of images.

The first one is a 2D convolutional network with a residual block (15) and a fully-connected layer, shown in Figure 3, referred to as the "2D stacked" model. Slices in a localizer are stacked on the channel dimension. Input to the network is a tensor whose height and width equal to that of each slice and channel length equal to the number of slices. The channel length of the inputs to this model must be constant, so all localizer stacks are zero padded to have a channel length equal to the maximum number of slices per stack.

The second one is a 3D fully convolutional network, shown in Figure 3, referred to as the "3D" model. This model can take inputs with varying number of slices if it is fully convolutional. However, it takes more runtime and memory than the 2D model.

**Shared 2D Feature Extractor with Attention**

To get the best of both networks above, we propose using a single-channel-input 2D CNN as the feature extractor shared among all slices, then regress from a linear combination of the extracted features. This architecture allows for a flexible stack size and has less parameters than both networks above.

Furthermore, to obtain an informative combination of slice features, we propose using an attention network to discriminatively weigh the features. Because a localizer stack contains up to 40 slices, many of them do not contain relevant anatomy for determining the ROI. A two-layer attention network is shared across all slices and trained implicitly within the whole end-to-end network. It takes as its input an output of the feature extractor and then outputs a scalar representing the importance of the corresponding slice. The resulting scalars from all slices are then passed through a SoftMax layer to get weights for the extracted features.

The weighted mean of the slice features has a fixed shape regardless of the input stack size. It is regressed to the final positional scalar outputs through a few more convolutional layers and a fully connected layer. The proposed architecture and training flow are illustrated in Figure 4.

All neural networks are trained with an $\ell_2$ loss on the boundary positions.

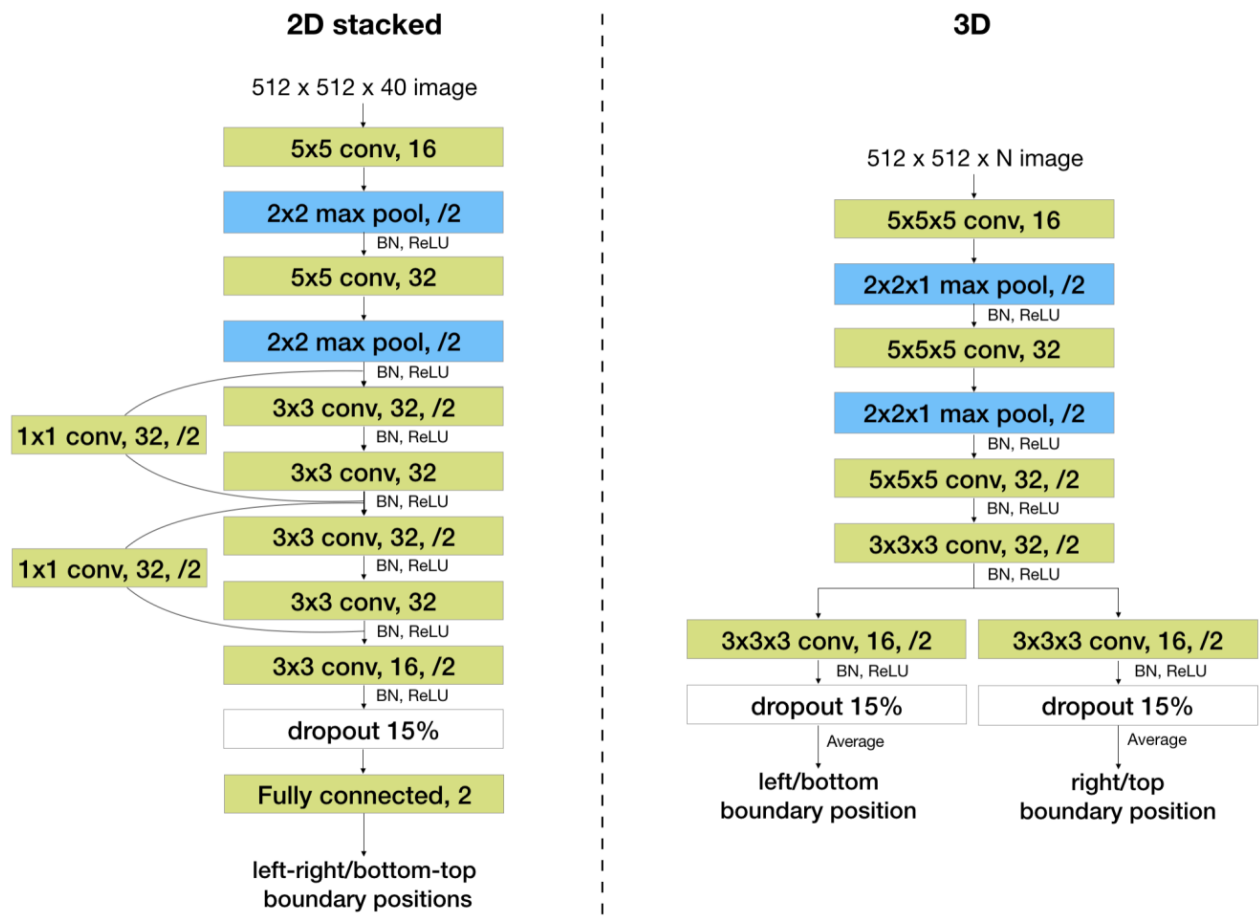

**Figure 3:** Two baseline network architectures. Network architectures of the 2D stacked baseline model (left) and the 3D baseline model (right). For example, "3x3 conv, 32, /2" represents a convolutional layer with 32 kernels of size 3x3 and using a stride of 2. BN stands for batch normalization. "Fully connected, 2" represents a fully connected layer with two output nodes.

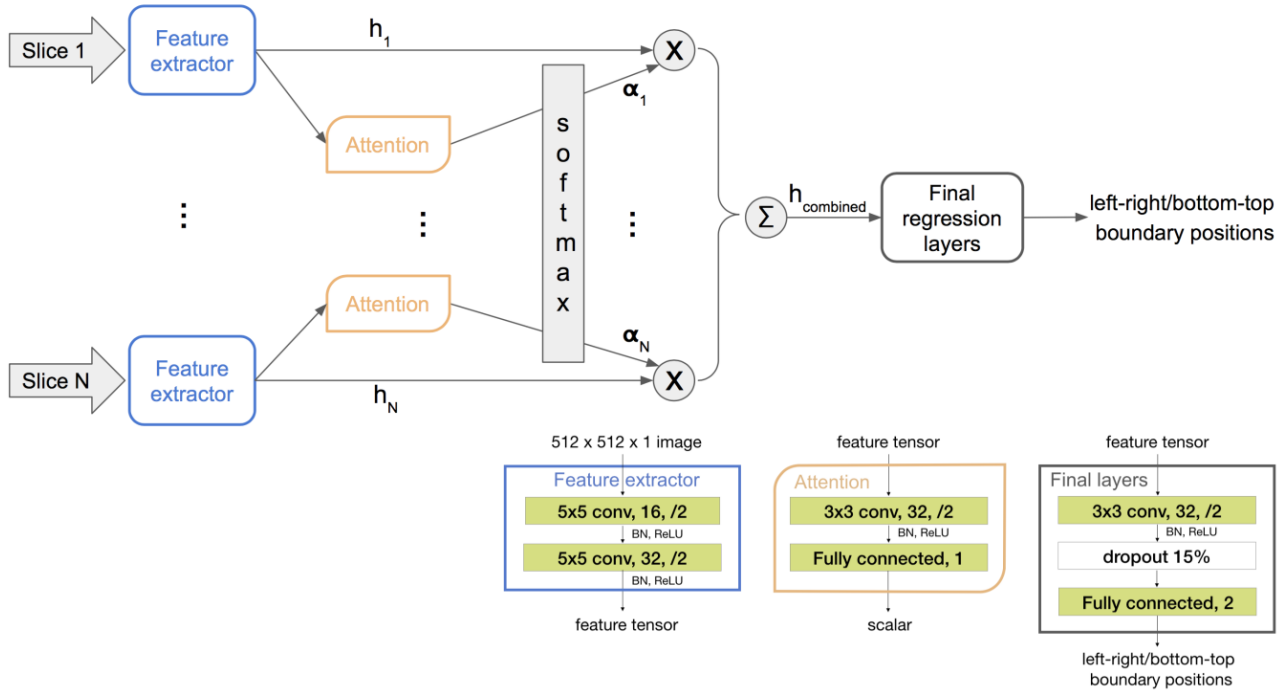

**Figure 4:** Network flow (top half) and architectures (bottom right corner) of the proposed framework. x represents scaling and Σ represents summation. α are scalars that sum to one. h are image features. The number following "Fully connected" represents the number of output nodes of that layer. There is only one instance of the feature extractor and the attention network, and they are shared across all slices.

**ROI to Smallest FOV**

The final goal of this work is to assign the smallest FOV that ensures no aliasing in the predicted ROI. We determine this FOV from a simplified model of the imaging object and the physics of MRI sampling.

First, we find the first and last row and column of an image where the pixel value sum surpasses a threshold. These four lines constitute a rectangular mask of the object in that slice. For simplicity, we assume that every pixel in this mask has a non-negligible signal, i.e., this mask represents the object.

In the phase encoding direction, two aliasing copies come from the original object shifted for the FOV length each way, as illustrated by Figure 5 (a). The smallest FOV should place the aliasing copy right outside the boundry of the ROI that is closer to that of the object mask. Let the width of the object mask be $y$, the distances between the ROI and FOV boundaries on two sides be $a$ and $b$, respectively, where $a \leq b$. Then the smallest FOV width is $y - a$. This FOV width leads to an alias-free region of width $y - 2a$ at the center of the object mask, regardless of the FOV position. Since $y - 2a \geq y - a - b$, the ROI width, there is a range of positions where the FOV box can be to include the ROI. We choose to put one of the FOV boundaries at the midpoint between the object mask and the ROI, $y - \frac{a}{2}$, to center the alias-free region, as shown in Figure 5 (b).

In the readout direction, aliasing is easily avoided due to a high sampling rate, and the FOV is set by the anti-aliasing filter. We set the FOV boundaries equal to those two of the ROI in this direction.

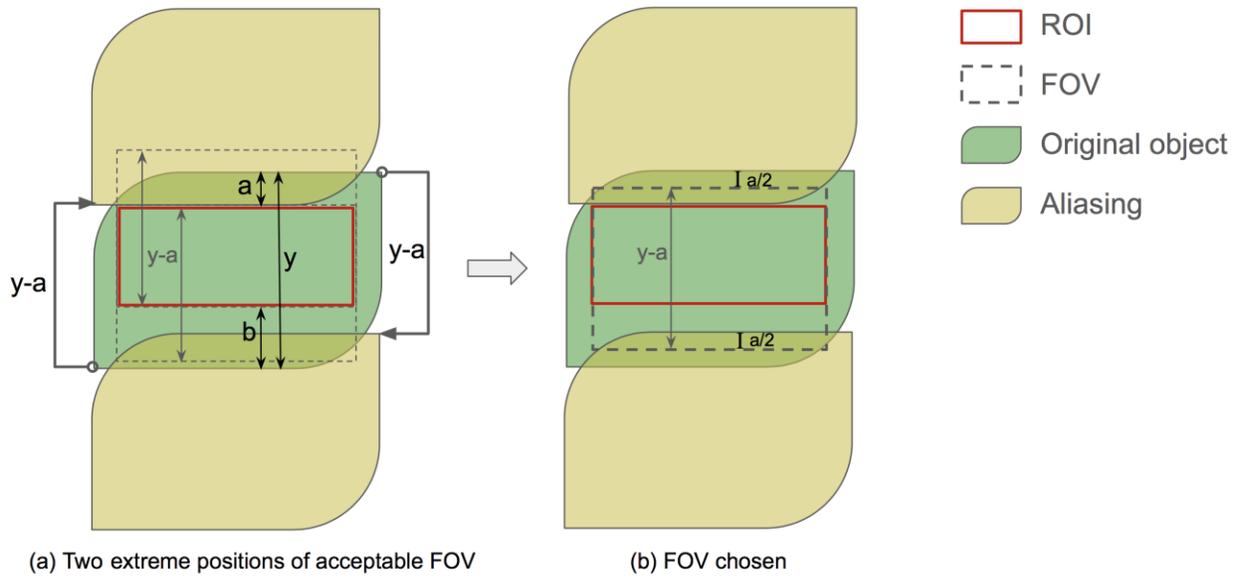

**Figure 5:** Determine the smallest FOV in the phase encoding direction (i.e., up-down direction in the figure) to achieve an aliasing free ROI. The object is aliased to integer numbers of FOV away from the original position, and two of the closest copies are shown in the figure. Given $a \leq b$, the smallest FOV width to keep the ROI alias free is $y - a$. The position of FOV does not affect the alias free region, but we need the ROI to be in the FOV. (a) shows two FOV at the extreme positions to fully include the ROI, and everything in between is acceptable. The position we use for this work is shown in (b).

Finally, the union of the FOV from all slices in the localizer is used for the upcoming scan.

**Evaluation Measures**

We use two quantitative metrics to evaluate the performance of the ROI prediction, given that we have ground truth labels for the ROI. We perform a qualitative reader study for the FOV assignment, which is the final output of the proposed framework.

Area intersection over union (IoU) and position error are used to measure the difference between two ROI. The position error is the distance between two boundary lines measured in pixels, and averaged across the four sides of an ROI box. Labels from Radiologist A are used as the ground-truth for the coronal test sets; labels from Radiologist B are used as the ground-truth for the axial test sets. ROI given by the models and the other radiologist is compared to these ground-truths.

For the qualitative reader study, the aliasing free region resulting from our model predicted FOV is shown to Radiologist A, and we ask if this aliasing free region is clinically acceptable.

We conduct two-tailed t-tests using SciPy 1.1.0 on the null hypothesis of the two methods having comparable position error or IoU. $P$ values of 0.05 and 0.01 are used for significance.

# Results

**Baseline Models**

ROI predicted by the standard 2D model on channel stacked inputs and the 3D convolutional model are evaluated by comparing to the labels given by a radiologist on four datasets: axial pelvic, coronal pelvic, axial abdominal, and coronal abdominal localizers. Each test dataset consists of 20 cases.

The boundary position error and IoU from the above comparison are shown in the first two columns of Table 1 and 2, respectively. The 2D model achieves average position errors ranging from 12.76 to 22.60 pixels out of 512 pixels and average IoU ranging from 0.693 to 0.846 on four datasets. The 3D model achieves average position errors ranging from 11.25 to 24.33 pixels out of 512 pixels and average IoU ranging from 0.701 to 0.852 on four datasets. Both baseline models are significantly ($P < 0.01$) worse than the proposed model and the radiologists. Both models perform better on the abdominal dataset than on the pelvic one. Neither of the two baseline models performs consistently better than the other. However, the 3D convolution model has four times longer inference time than the 2D convolution based models.

**Shared 2D Feature Extractor with Attention**

The shared 2D feature extractor and attention model is evaluated by comparing its ROI prediction to the labels given by a radiologist on the same four test datasets as above. The difference between this model and a radiologist, shown in the third column of Table 1 and 2, is then compared with the difference between two radiologists, shown in the fourth column of Table 1 and 2.

Table 1: Comparison of area intersection over union (IoU) averaged over cases.

| Dataset | 2D stacked | 3D | Intra-stack attention | Radiologist |
|---|---|---|---|---|
| Pelvis (axial) | 0.751 ± 0.169[†] | 0.701 ± 0.183[‡] | 0.863 ± 0.106 | **0.874 ± 0.048** |
| Pelvis (coronal) | 0.693 ± 0.186[‡] | 0.722 ± 0.122[‡] | **0.835 ± 0.056** | 0.815 ± 0.074 |
| Abdomen (axial) | 0.846 ± 0.055[‡] | 0.852 ± 0.051[‡] | 0.896 ± 0.033 | **0.916 ± 0.022**[†] |
| Abdomen (coronal) | 0.833 ± 0.062 | 0.838 ± 0.066[†] | 0.884 ± 0.060 | **0.877 ± 0.086** |

*Note.*— Three models are compared with the inter-rater variance between two radiologists across four datasets with 20 cases each. The mean ± standard deviation are reported, with significance levels noted by:

[†] $0.01 \leq P < 0.05$ in comparison to intra-stack attention (ours),
[‡] $P < 0.01$ in comparison to intra-stack attention.

Table 2: Comparison of boundary position error averaged over four sides then over cases.

| Dataset | 2D stacked | 3D | Intra-stack attention | Radiologist |
|---|---|---|---|---|
| Pelvis (axial) | 17.13 ± 12.31[†] | 22.98 ± 17.30[‡] | 10.26 ± 8.61 | **7.02 ± 3.71** |
| Pelvis (coronal) | 22.60 ± 16.15[‡] | 24.33 ± 20.48[‡] | **11.61 ± 4.18** | 14.32 ± 5.72 |
| Abdomen (axial) | 12.76 ± 9.49[‡] | 11.25 ± 7.96[‡] | 6.15 ± 1.97 | **5.28 ± 1.52** |
| Abdomen (coronal) | 13.47 ± 8.83[†] | 12.96 ± 8.55[†] | **8.23 ± 4.05** | 9.46 ± 8.25 |

*Note.*— Three models are compared with the inter-rater variance between two radiologists across four datasets with 20 cases each. The mean ± standard deviation are reported, with significance levels noted by:

[†] $0.01 \leq P < 0.05$ in comparison to intra-stack attention (ours),
[‡] $P < 0.01$ in comparison to intra-stack attention.

The proposed model achieves average boundary position errors ranging from 6.15 to 11.61 pixels out of 512 pixels and average IoU ranging from 0.835 to 0.874 on four datasets. This performance is significantly better than both baseline models ($P < 0.01$ on ten and $P < 0.05$ on the other six out of sixteen comparisons), and comparable to that of a radiologist ($P > 0.12$ on seven out of eight comparisons).

Figure 6 and Figure 7 provide visual examples of the predicted ROI compared with the labeled ROI on pelvic and abdominal localizers, respectively. While the ROI is based on the whole stack, one representative slice is used for illustration.

**FOV Acceptability**

We present the resulting alias-free region when using the FOV given by our framework to a radiologist and ask whether this end result is clinically acceptable, i.e., ROI is within the alias-free region and not too much smaller than it. The radiologist is given three ratings to choose from: yes, almost, and no. 40 pelvic cases and 40 abdominal cases are included in this study. 69 cases get the rating "yes", 9 cases get the rating "almost", and 2 cases get the rating "no".

We calculate the 80% confidence interval for proportions for the counts above. The frequency interval of the "yes" rating is 0.80 to 0.91, and the frequency interval of the "yes" and "almost" ratings are 0.93 to 0.99.

**Inference Time and Implementation**

The average inference time for one stack of localizer images is 0.12, 0.44, and 0.12 seconds for the 2D stacked baseline, 3D baseline, and 2D intra-stack attention model, respectively. A batch size of one, and one NVIDIA GeForce GTX 1080 GPU is used for the inference. All models are trained from random parameter initialization. The models are implemented in TensorFlow and Keras. A GitHub link to the code will be given in the publication version.

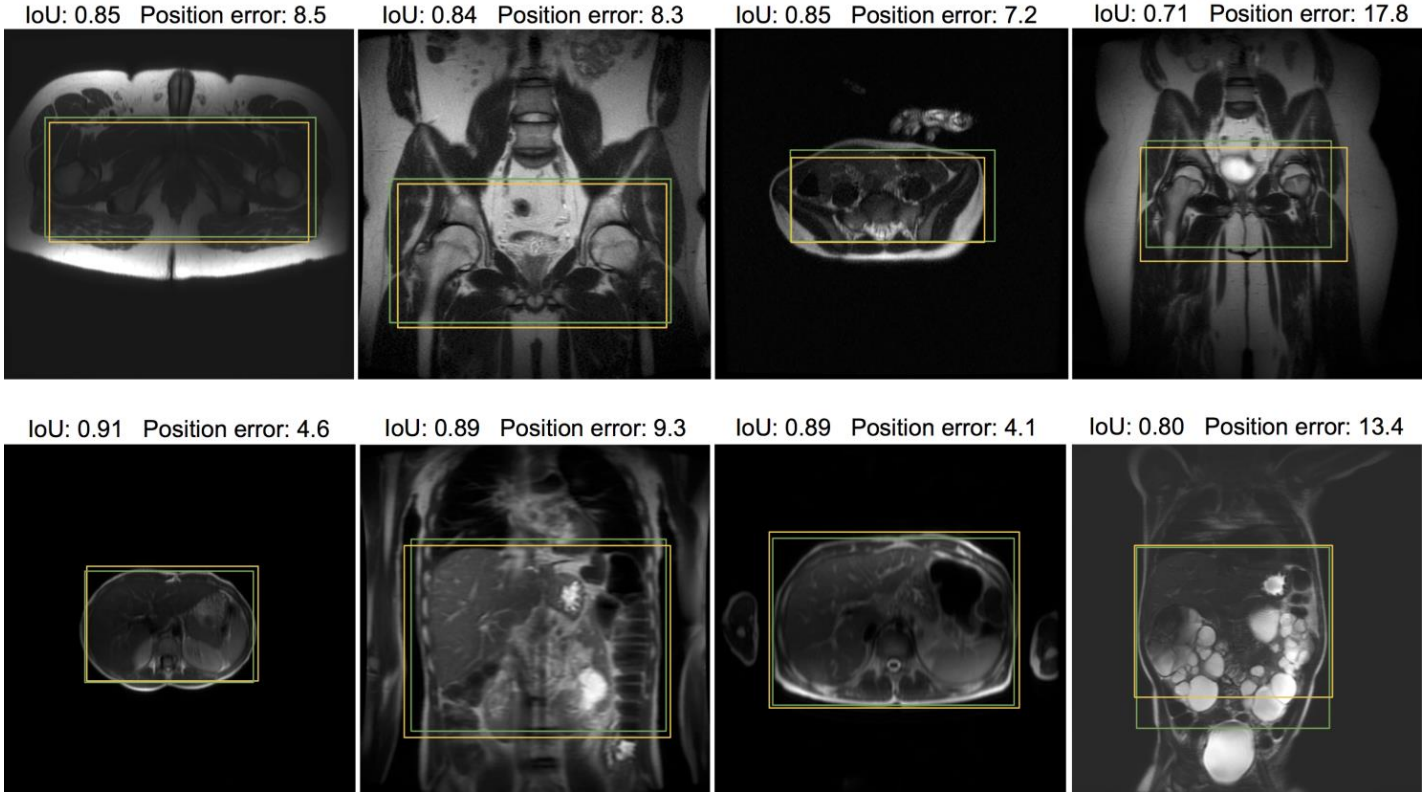

**Figure 6:** Samples from the test dataset annotated with intra-stack attention model predicted ROI (yellow) and the ROI label given by radiologists (green). On the first row are pelvic localizers, and on the second row are abdominal localizers. Position error and IoU for the model prediction in each example are included to visualize the quantitative results. Note that only one representative slice of a stack is shown, and neither ROI is determined merely from this slice. The bottom right example highlights an extreme case of polycystic kidney disease.

## Discussion

We find a CNN-based model with intra-stack parameter sharing and attention suitable for automating the FOV prescription step in MR scans. On the FOV, prescription given by the proposed model achieves a 92% acceptance rate (cases rated as "almost" acceptable are counted as halves) for clinical scans. On the ROI, the proposed model achieves an average IoU of 0.849 and average position error of 10.93 out of 512 pixels on 40 pelvic scans. It achieves an average IoU of 0.885 and average position error of 7.19 out of 512 pixels on 40 abdominal scans. This performance is significantly better than standard CNN models without attention or parameter sharing, and is not significantly inferior to the variance between two radiologists. The model predicted FOV is significantly more compact than the manually prescribed one by MRI technologists, which despite being conservatively large, sometimes misses parts of interests.

We notice that the objects in abdominal localizers appear more zoomed out than those in the pelvic localizer. And the abdominal ROI takes up a larger portion of the whole object shown in the localizers than the pelvic ROI. This may be one reason that our framework performs slightly better on the abdominal dataset than the pelvic dataset. This highlights that the technique used to obtain the localizers may be critical to performance.

There is no existing work on deep learning based end-to-end FOV box prediction to our knowledge. Our framework is at least 100 times faster than existing non-deep-learning methods.

Its inference time is 0.12 seconds per stack of images, on a GPU, and 0.17 seconds per stack on a CPU, compared to 10 seconds or 1 minute from previous works (5-6). The existing works (1-7) are done on brain or knee images and cannot be adapted to our pelvic and abdominal images without additional information. Therefore, in this work, the performance comparison is only performed across deep learning models.

Besides the target application of this work, automatic scan prescriptions, the ROI prediction network can be useful for image quality assessment when the FOV is larger than the ROI. Instead of assessing the whole image (16), assessing only the ROI gives more relevant and accurate evaluation results. This difference in region of assessment is particularly influential with respect to localized artifacts that degrade image quality.

One limitation of the presented framework is that it does not support oblique ROI and FOV prediction. Some objects in the localizer images are rotated, in which case the most efficient FOV is aligned with the object and not aligned with the localizer image. Adding support on predicting a rotated box is the next step for future work. The majority of the current framework holds, except that the neural network then needs to output five or six scalars defining a rotated rectangle instead of four scalars defining a rectangle along the coordinate axes.

We have presented a framework for ROI and FOV prediction on pelvic and abdominal MR scans. The framework utilizes a neural network model for ROI prediction from a stack of localizers, followed by a thresholding step and a MR physics derived algebraic conversion to FOV. We have proposed an architecture consisting of slice-wise shared CNN and attention mechanism for the bounding box regression model, and compared its performance with standard CNN models and a radiologist. The framework has been examined on a set of cases with large patient size/shape variation and abnormalities. It can be applied clinically to reduce the time gap between series of scans, and to get a more efficient and precise FOV for diagnostic scans.